\documentclass{bmcart}

\usepackage{amsthm,amsmath}
\RequirePackage{natbib}
\usepackage[utf8]{inputenc} 
\usepackage{url}
\usepackage{graphicx}
\usepackage{booktabs}
\usepackage{multicol}
\usepackage{multirow}
\usepackage{lscape}
\usepackage{afterpage}
\usepackage{comment}
\usepackage{soul}
\usepackage{todonotes}

\startlocaldefs
\endlocaldefs

\begin{document}

\begin{frontmatter}

\begin{fmbox}
\dochead{Methodology}


\title{A continuous-time state-space model for rapid quality-control of Argos locations from animal-borne tags}


\author[
   addressref={aff1},                   
   email={ian.jonsen@mq.edu.au}   
]{\inits{IDJ}\fnm{Ian D.} \snm{Jonsen}}
\author[
   addressref={aff2},
   email={toby.patterson@csiro.au}
]{\inits{TAP}\fnm{Toby A.} \snm{Patterson}}
\author[
   addressref={aff3},
   email={costa@ucsc.edu}
]{\inits{DPC}\fnm{Daniel P.} \snm{Costa}}
\author[
   addressref={aff4},
   email={p.doherty@exeter.ac.uk}
]{\inits{PDD}\fnm{Philip D.} \snm{Doherty}}
\author[
   addressref={aff4},
   email={b.j.godley@exeter.ac.uk}
]{\inits{BJG}\fnm{Brendan J.} \snm{Godley}}
\author[
   addressref={aff6 },
   email={wjg5@st-andrews.ac.uk}
]{\inits{WJG}\fnm{W. James} \snm{Grecian}}
\author[
   addressref={aff7},
   email={guinet@cebc.cnrs.fr}
]{\inits{CG}\fnm{Christophe} \snm{Guinet}}
\author[
   addressref={aff2},
   email={xavier.hoenner@csiro.au}
]{\inits{XH}\fnm{Xavier} \snm{Hoenner}}
\author[
   addressref={aff3},
   email={sarah.stachura@gmail.com}
]{\inits{SSK}\fnm{Sarah S.} \snm{Kienle}}
\author[
   addressref={aff3},
   email={patrick.robinson@ucsc.edu}
]{\inits{PWR}\fnm{Patrick W.} \snm{Robinson}}
\author[
   addressref={aff4},
   email={s.c.votier@exeter.ac.uk}
]{\inits{SCV}\fnm{Stephen C.} \snm{Votier}}
\author[
    addressref={aff8},
    email={scott.whiting@dbca.wa.gov.au}
]{\inits{SW}\fnm{Scott} \snm{Whiting}}
\author[
   addressref={aff4},
   email={m.j.witt@exeter.ac.uk}
]{\inits{MJW}\fnm{Matthew J.} \snm{Witt}}
\author[
   addressref={aff9},
   email={mark.hindell@utas.edu.au}
]{\inits{MAH}\fnm{Mark A.} \snm{Hindell}}
\author[
   addressref={aff1},
   email={robert.harcourt@mq.edu.au}
]{\inits{RGH}\fnm{Robert G.} \snm{Harcourt}}
\author[
   addressref={aff10},
   email={clive.mcmahon@utas.edu.au}
]{\inits{CRM}\fnm{Clive R.} \snm{McMahon}}

\address[id=aff1]{
  \orgname{Dept of Biological Sciences, Macquarie University}, 
  \city{Sydney},                              
  \cny{Australia}                                    
}
\address[id=aff2]{%
  \orgname{CSIRO Oceans and Atmosphere},
  \city{Hobart},
  \cny{Australia}
}
\address[id=aff3]{%
  \orgname{Department of Ecology and Evolutionary Biology, University of California Santa Cruz},
  \city{Santa Cruz},
  \cny{USA}
}
\address[id=aff4]{%
  \orgname{Environment and Sustainability Institute, University of Exeter},
  \city{Penryn},
  \cny{UK}
}
\address[id=aff6]{%
  \orgname{Sea Mammal Research Unit, Scottish Oceans Institute, University of St Andrews},
  \city{St Andrews},
  \cny{UK}
}
\address[id=aff7]{%
  \orgname{CNRS},
  \city{Chize},
  \cny{France}
}
\address[id=aff8]{%
    \orgname{Department of Biodiversity, Conservation and Attractions, Government of Western Australia},
    \city{Kensington},
    \cny{Australia}
    }
\address[id=aff9]{%
  \orgname{Institute for Marine and Antarctic Studies, University of Tasmania},
  \city{Hobart},
  \cny{Australia}
}
\address[id=aff10]{%
  \orgname{Sydney Institute of Marine Science},
  \city{Mosman},
  \cny{Australia}
}


\begin{artnotes}
\end{artnotes}

\end{fmbox}


\begin{abstractbox}

\begin{abstract} 
\parttitle{Background} 
State-space models are important tools for quality control and analysis of error-prone animal movement data. The near real-time (within 24 h) capability of the Argos satellite system can aid dynamic ocean management of human activities by informing when animals enter wind farms, shipping lanes, and other intensive use zones. This capability also facilitates the use of ocean observations from animal-borne sensors in operational ocean forecasting models. Such near real-time data provision requires rapid, reliable quality control to deal with error-prone Argos locations.

\parttitle{Methods} 
We formulate a continuous-time state-space model to filter the three types of Argos location data (Least-Squares, Kalman filter, and Kalman smoother), accounting for irregular timing of observations. Our model is deliberately simple to ensure speed and reliability for automated, near real-time quality control of Argos location data. We validate the model by fitting to Argos locations collected from 61 individuals across 7 marine vertebrates and compare model-estimated locations to contemporaneous GPS locations. We then test assumptions that Argos Kalman filter/smoother error ellipses are unbiased, and that Argos Kalman smoother location accuracy cannot be improved by subsequent state-space modelling.

\parttitle{Results} 
Estimation accuracy varied among species with median Root Mean Squared Errors usually $<$ 5 km, and decreased with increasing data sampling rate and precision of Argos locations. Including a model parameter to inflate Argos error ellipse sizes in the north - south direction resulted in more accurate location estimates. Finally, in some cases the model appreciably improved the accuracy of the Argos Kalman smoother locations, which should not be possible if the smoother is using all available information.

\parttitle{Conclusions} 
Our model provides quality controlled locations from Argos Least-Squares or Kalman filter data with accuracy similar to or marginally better than Argos Kalman smoother data that are only available via fee-based reprocessing. Simplicity and ease of use make the model suitable both for automated quality control of near real-time Argos data and for manual use by researchers working with historical Argos data.
\end{abstract}




\end{abstractbox}
%

\end{frontmatter}




\section*{Background}
State-space models have emerged as important tools both for quality control and ecological analysis of error-prone animal movement data \cite{Jonsen:2005,Johnson:2008,Patterson:2008,Albertsen:2015,AugerMethe:2017}. Analysis of these data with discrete-time models is simple in principle, breaking down animal movement into a series of discrete steps that occur on some fixed time interval (e.g., \cite{Jonsen:2005,McClintock:2012}). Yet animal movement is a process that unfolds continuously through time, usually absent of clear breaks that could delineate discrete steps. We merely measure the movements from locations obtained over discrete, often irregular intervals in time. In this sense, a continuous-time model can more naturally handle temporally irregular observations while mimicking the true underlying continuous movement process(es) \cite{Johnson:2008,McClintock:2014}.

In the marine realm, air-breathing animal locations are typically measured by satellite-linked electronic tags at irregular time intervals dictated by a combination of satellite availability and an animal's surface behaviour. The Argos satellite telemetry system is one of the most common platforms used to track animals at sea, with over 40,000 individuals tracked since 2007 (S. Baudel, pers. comm.). In this system, transmissions from electronic tags are received by one of several polar-orbiting satellites as they pass overhead, and the Doppler shift in transmission frequency along with other information is used to geolocate the tags \cite{CLS:2016}. The polar orbits of Argos satellites result in more dense coverage and potentially higher temporal, resolution data closer to the poles than at the equator. From inception in 1978 to 2011, CLS (Collecte Localisation Satellites) has used a Least-Squares algorithm to geolocate the tag transmissions. This approach does not quantify location uncertainty but rather provides location quality classes based on information including the number of transmissions received \cite{CLS:2016}. 

State-space models developed for Argos Least-Squares locations have relied on independent, ground-truth data (e.g., \cite{Vincent:2002}) to quantify location uncertainty for each of the location quality classes \cite{Jonsen:2005,Johnson:2008}. However, independently quantified uncertainties, based on a single or small number of data sets, are unlikely to be appropriate for all species in all locations. For example, Lowther et al. \cite{Lowther:2015} found that modifications to assumed Least-Square error variances can influence the accuracy of locations predicted by different state-space models.  

In 2011, CLS replaced their Least-Squares algorithm with a state-space model, based on a multiple model Kalman filter algorithm, to estimate locations and their uncertainty \cite{Lopez:2014}. This approach provides more location estimates, each with a corresponding estimated error ellipse, and with greater accuracy compared to the original Least-Squares method. These locations are provided in near real-time; here defined as within 24 h of occurrence. However, CLS also provides an extra service that uses a fixed-interval Kalman smoother to further improve location accuracy from the original Kalman filter-based location estimates \cite{Lopez:2015}. Whereas the Kalman filter employs a one-step recursion to estimate locations based only on the current and previous observations, the Kalman smoother uses a two-pass approach, first employing the Kalman filter and then employing a backwards smooth of the data \cite{Rauch:1965}. In this sense, the Kalman smoother uses information from the entire animal track to estimate locations and their uncertainty. This results in more accurate location estimates than the Kalman filter alone \cite{Lopez:2015}. Such smoother-based location estimates are theoretically optimal given the available data, and it should not be possible to improve on them if uncertainty is characterised and propagated accurately (e.g., \cite{McClintock:2015}). Currently, CLS does not provide Kalman smoother-based locations in near real-time, they can only be obtained with reprocessing, for an additional fee, after a tag deployment ends. 

Traditional use of animal tracking data has required neither near real-time data provision nor rapid modelling tools for quality control or ecological analysis. However, real-time management of at-risk species' mortality from interactions with human activities such as offshore wind farms, fisheries and shipping increasingly relies on animal telemetry data \cite{Maxwell:2015,Hazen:2018,Pirotta:2019}. Dynamic ocean management applied at high spatial and temporal resolutions can increase the efficiency and efficacy of measures to reduce mortality \cite{Dunn:2016}, placing an onus on rapidly available, high-resolution data. Similarly, the utility of animal-borne sensors for ocean observing \cite{Treasure:2017,Harcourt:2019} as part of the Global Ocean Observing System has spurred coordinated animal telemetry programs, such as the Australian Integrated Marine Observing System's Animal Tracking Facility (IMOS ATF\footnote{\url{http://imos.org.au/facilities/animaltracking}}) and the U.S. Integrated Ocean Observing System's Animal Telemetry Network (IOOS ATN\footnote{\url{https://ioos.noaa.gov/project/atn}}). These programs aim to provide near real-time ocean measurements via the World Meteorological Organization's Global Telecommunication System for assimilation in operational ocean and atmospheric forecast models. In all these cases, near real-time telemetry data provision requires rapid and therefore automated, reliable quality control processes, including the error-prone Argos location data that are essential for understanding animal movements and distribution, and for providing geospatial context to ocean measurements.

Here we present a continuous-time state-space model for rapid filtering of any Argos location data. This model is now used as part of the IMOS ATF's quality control/quality assurance process for animal-borne ocean observations. To facilitate fast automation, we trade off realism - the ability to explain complex movement processes - for reliability by using a simple continuous-time random walk on velocity with a single variance parameter. We evaluate the model by: 1) comparing fits to all three Argos location types from the same individuals; 2) assessing accuracy of model-estimated locations against contemporaneous GPS locations; 3) assessing how a model assumption about Argos error ellipses influences estimation accuracy; 4) comparing the accuracy of modelled and un-modelled Kalman Smoother locations.

\section*{Methods}
\subsection*{A continuous-time state-space model for animal telemetry data}
We model animal movement as a continuous-time random walk on velocity ${\bf v}_t$ in two coordinate axes:
\begin{equation}\label{eq:vel}
  {\bf v}_t = {\bf v}_{t-\Delta} + {\Sigma}_{\Delta}
\end{equation}
where $\Delta$ is the time increment and ${\Sigma}_\Delta$ is a zero-mean, bi-variate Gaussian random variable with variance $2\textrm{D}\Delta$. The parameter $\textrm{D}$ is a 1-d diffusion coefficient accounting for variability in velocity, which increases with the time interval $\Delta$. Noting that locations ${\bf x}$ are the summed velocities, given a starting location, the following equation describes a simple process model subject to variable time increments:

\begin{equation}\label{eq:loc}
  {\bf x}_i = {\bf x}_{i-1} + {\bf v}_i\Delta_i
\end{equation}
where the subscript $i$ indexes time $t_i$, ${\bf x}_i$ is the true location of the animal at time $t_i$ and ${\bf v}_i\Delta_i$ is the displacement (velocity x elapsed time) between ${\bf x}_{i-1}$ and ${\bf x}_i$. To simplify the model, we assume that the velocity random walk variances $2\textrm{D}\Delta_i$ are equal on the two axes but they could also be assumed to vary independently \cite{Johnson:2008}. Correlation in movements arises from allowing the locations to be the sum of the velocities. 

We couple this process model to a generally applicable measurement model that describes how the error-prone and possibly irregularly-timed observed locations ${\bf y}_i$ map onto the corresponding true location states ${\bf x}_i$:
\begin{equation}\label{eq:meas}
  {\bf y}_i = {\bf x}_i + {\epsilon}_i; \hspace{1cm} {\epsilon}_i \sim {\textrm N}(0, { \Omega_i})
\end{equation}
\noindent where ${\bf y}_i$ the location observed at time $t_i$ corresponding to ${\bf x}_i$ 
, and ${\Omega_i}$ is the measurement error variance-covariance matrix that can be structured to suit different types of location data. Below, we focus on modifications to accommodate different Argos location types, but other location data (e.g., processed light-level geolocations) could also be considered in this framework.

\subsubsection*{Argos Least-Squares data}
Locations measured using CLS' older Least-Squares (LS) approach \cite{CLS:2016} are associated with location quality class designations: 3, 2, 1 0, A, B, and Z. These classes are the only contemporaneous information about location quality and provide only a relative index of measurement uncertainty \cite{Jonsen:2005}. We use the class information, along with independent estimates of their associated standard errors from Argos transmitters deployed on seals held captive at a known location \cite{Vincent:2002}, to construct the following variance-covariance matrix: 

\begin{equation}\label{eq:covarLS}
  \Omega_i = 
  \begin{bmatrix} 
    \tau_x^2 K_{x,i}^2 & \rho \tau_x K_{x,i} \tau_y K_{y,i}\\
    \rho \tau_x K_{x,i} \tau_y K_{y,i} & \tau_y^2 K_{y,i}^2 
  \end{bmatrix}
  \end{equation}

\noindent where $\tau_x^2$ and $\tau_y^2$ are the overall measurement error variances on the two coordinate axes, $K_{x,i}$ and $K_{y,i}$ are error weighting factors that scale the ${\bf \tau}$'s appropriately for the Argos location quality class associated with the $i$th observation. The $\tau$'s are estimated during model fitting and the error weighting factors are the standard error ratios between the best quality class, 3, and each other class (2, 1, 0, A, B, Z).

\subsubsection*{Argos Kalman filter and Kalman smoother data}
Locations measured using CLS' Kalman filter (KF) or Kalman smoother (KS) algorithms have their estimated uncertainties provided to users as error ellipses \cite{Lopez:2014}. Ellipses are defined by three variables: semi-major axis, semi-minor axis and semi-major axis orientation from north. Building on McClintock et al.\cite{McClintock:2015}, the error variance-covariance matrix is: 
\begin{equation}\label{eq:covarKF}
  {\Omega}_i = 
    \begin{bmatrix}
      \tau_{x,i}^2 & \tau_{xy,i} \\
      \tau_{xy,i} & \tau_{y,i}^2
    \end{bmatrix}
\end{equation}

\noindent with the elements being derived from the Argos error ellipse components:
\begin{align}\label{eq:elps1}
  \tau_{x,i}^2 = \left( \frac{M_i}{\sqrt{ 2}} \right)^2 \sin^2 c_i + \left( \frac{m_i\psi}{\sqrt{ 2}} \right)^2 \cos^2 c_i\\
  \tau_{y,i}^2 = \left( \frac{M_i}{\sqrt{ 2}} \right)^2 \cos^2 c_i + \left( \frac{m_i\psi}{\sqrt{ 2}} \right)^2 \sin^2 c_i
\end{align}
and
\begin{equation}\label{eq:elps2}
  \tau_{xy,i} = \left(\frac{M^2_i - m_i^2\psi^2}{2} \right) \cos c_i \sin c_i
\end{equation}
\noindent where $M_i$ is the ellipse semi-major axis length of the $i$-th observation, $m_i$ is the semi-minor axis length and $c_i$ is the semi-major axis orientation \cite{Lopez:2014,McClintock:2015}. 

McClintock et al. \cite{McClintock:2015} used a bivariate $t$-distribution, with variance-covariance defined by the Argos error ellipses, in their measurement model to account for occasional outlier observations (i.e., where error ellipses underestimate the true measurement uncertainty). Here we chose to identify and remove outlier locations using a travel rate filter \cite{Freitas:2008} prior to fitting the state-space model, as per \cite{Johnson:2008,Patterson:2010}. Additionally, we included the parameter $\psi$ to account for possible consistent under estimation of the Kalman filter (\& smoother)-derived location uncertainty (Figure \ref{fig:EE}). $\psi$ re-scales all ellipse semi-minor axes $m_i$, where estimated values $> 1$ inflate the uncertainty region around measured locations by lengthening ellipse semi-minor axes. 

In all cases, we project the ${\bf y}_i$'s from geographic coordinates (lon, lat) onto a Cartesian plane prior to modelling, using the WGS84 World Mercator projection (EPSG 3395). To facilitate optimization, all planar coordinates and their uncertainty estimates, where available, are converted from m to km.

\subsection*{Estimation}
We used the R package TMB (Template Model Builder, \cite{Kristensen:2016}) to fit the state-space model, using maximum likelihood to estimate model parameters and the Laplace approximation to rapidly estimate the random effects - the unobserved location and velocity states, ${\bf x}$ and ${\bf v}$ \cite{AugerMethe:2017,Jonsen:2019}. Using this estimation approach, uncertainty in ${\bf x}$ and ${\bf v}$ estimates are obtained using a generalised delta method (see \cite{Kristensen:2016} for details). All model and associated general data preparation code are available in the \texttt{foieGras} R package \cite{foieGras}. The latest version can be downloaded from the lead author's GitHub site (\url{https://github.com/ianjonsen/foieGras}).

\subsection*{Data and pre-processing}
We model all three types of Argos satellite location data: LS, KF, and KS. The data are comprised of four pinnipeds, one seabird and two sea turtle species (Table \ref{tab:sppdat}); with deployment locations ranging between polar, temperate, and tropical marine regions (Figure \ref{fig:sppmap}). The number of individual data sets by species and data type range from 6 to 13 with all having locations measured by GPS and at least one Argos type (Table \ref{tab:sppdat}). All data collected after 2008 were reprocessed by CLS to obtain the three Argos data types (4 species; Table \ref{tab:sppdat}). 

We used an automated pre-filtering step to identify outlier observations to be ignored by the state-space model. This pre-filtering used the \texttt{argosfilter} R package \cite{Freitas:2008} to identify locations implying travel rates $>$ 3 ms\textsuperscript{-1} for all pinnipeds and sea turtles and travel rates $>$ 17 ms\textsuperscript{-1} for northern gannets. These speed thresholds represent conservative upper limits of travel for these species and are intended to identify only the extreme outlier observations. This resulted in $< 30\%$ of Least-Squares, $< 15\%$ of Kalman filter, and $< 10\%$ of Kalman smoother data being removed. The proportion of data removed by pre-filtering is considerably less than those associated with optimal speed thresholds for other species (e.g.,  \cite{Patterson:2010}).

\subsection*{Empirical validation}
We examined the accuracy of model-predicted locations, assuming GPS data represent truth. Although GPS data have higher spatial accuracy and precision, and typically have higher sampling rates than Argos data, they are nonetheless discrete measurements of a continuous-time process. As a consequence, they are also likely to misrepresent animals' true movement paths but to a far smaller extent (10's of m; \cite{Bryant:2007}) than Argos data. 

For all validations presented, we compared GPS locations to model-fitted locations (hereafter model-estimated locations), which are location states estimated at the times of the Argos-measured locations. By focusing on model-estimated locations and not predicted locations that occur at regular time intervals, we reduce the degree to which model accuracy is confounded with data sampling rates that are known to vary across species and Argos data types (see Discussion). 

We compared model-estimated locations from fits to all three Argos data types, where available, with GPS data. In all cases, the times of GPS observations do not match the times of Argos observations or the corresponding model-estimated locations. To account for this mismatch, we initially considered three approaches for comparing between GPS and modelled locations. First, using a linear interpolation of GPS locations to model-estimated location times \cite{Silva:2014}. Second, using the temporally closest GPS observation if any occurred within $\pm$ 10 min. Third, using the model to predict locations at the GPS observation times. In several cases, it was not feasible to predict model locations for each GPS observation time as the typically higher frequency of GPS observations resulted either in implausible artefacts in the model fits to the Argos data or in convergence failures of the optimiser used to fit the model. For these reasons, we chose not to consider this approach further.

Fitting the state-space model with a fixed 2-h prediction interval resulted in optimiser convergence for all individual tracks. For each individual track, we summarized the deviations between model-estimated locations and either the linearly interpolated GPS locations or the temporally matched GPS locations by taking the root mean of the squared distances (RMSD in km) between all pairs of locations and comparing distributions of individual RMSD values among species. We report results of comparisons with the linearly interpolated GPS locations here and comparisons with the temporally matched GPS locations in Supplementary Information. We discuss implications of using each of these approaches.

\subsubsection*{Potential under-representation of Argos KF/KS location uncertainty}
Our default model accounts for a perceived under-estimation of the size of CLS' Kalman filter and Kalman smoother error ellipses (Figs. \ref{fig:EE} and \ref{fig:SellipseHB} - \ref{fig:SellipseNG}) by including the parameter $\psi$ (Eqns. \ref{eq:elps1}, \ref{eq:elps2}). Although uncertainty is expected to be lower in the general North - South plane due to the polar orbits of the Argos satellites \cite{Lopez:2014}, the frequent compression of error ellipses in this plane (semi-minor axis; e.g., Fig. \ref{fig:EE}b) seems extreme. Values of $\psi > 1$ inflate the semi-minor axis, increasing the uncertainty region around Argos KF/KS observations and could allow the model to more appropriately smooth the data. It is unclear how much the parameter actually improves the accuracy of estimated tracks versus yielding a less accurate over-smoothing of the data. To assess this, we evaluated the influence of the $\psi$ parameter on the accuracy of model-estimated locations by comparing RMSD values from models with and without the $\psi$ parameter. To simplify the results, we pooled RMSD values across species and assessed the $\log_{e}$ difference in RMSD (denoted as $\log \Delta$ RMSD), which approximates \% difference on the linear scale \cite{Tornqvist:1985}.

\subsubsection*{Argos KS location accuracy}
The CLS Kalman smoother locations have greater spatial accuracy and precision than Least-Squares or Kalman filter data \cite{Lopez:2015}. In principle, it should not be possible to improve the accuracy of KS-based locations with subsequent modelling because they are theoretically optimal estimates, using all available data. It does seem reasonable, however, to question whether this is actually the case. We evaluated this by comparing $\log \Delta$ RMSD derived from GPS and KS locations to those derived from GPS and estimates from the state-space model fit to the KS locations. In both cases, we apply the same pre-filtering to identify and remove outlier locations, though these outliers should not be present in KS-based locations.

\section*{Results}
\subsection*{State-space model fits to the 3 Argos data types}
We fit the state-space model to the four species with all three Argos data  (Table \ref{tab:sppdat}), and present fits with a 2-h prediction time interval. Model fits to hawksbill turtle and southern elephant seal data show a consistent increase in spatial resolution and decrease in estimation uncertainty of the predicted tracks across the three Argos data types (top to bottom; (Fig. \ref{fig:sppfits} a,e,i and b,f,j, respectively). This effect is due to an increase in the number of observations from least-squares to Kalman filter data, and to a shrinking of the error ellipses (measurement uncertainty), by nearly half, from Kalman filter to Kalman smoother data (Table \ref{tab:dataerr}). Model fits to leopard seal and northern gannet data do not show any clear differences in resolution or estimation uncertainty across the Argos data types (Fig. \ref{fig:sppfits} c,g,k and d,h,l, respectively). This appears due to smaller differences in the number of observations for Least-Squares versus Kalman filter data, arising from lower proportions of class A and B locations, relative to hawksbill turtles and southern elephant seals (Table \ref{tab:dataerr}). The lower proportions of class A and B locations for leopard seals and northern gannets are likely due to the large amount of time they spend at or above the ocean surface. Additionally, northern gannets had, on average, far larger error ellipses than the other species (Table \ref{tab:dataerr}). The uncertainty of their state-space model-predicted locations was consequently larger, regardless of Argos data type (light blue 95\% confidence ellipses in Fig. \ref{fig:sppfits} d,h,l).

\subsection*{Validation with GPS data}
Total sequential processing time for all 129 Argos data sets (Table \ref{tab:sppdat}) was 13.43 min, an average of 6.25 s per data set. This included both the pre-filter algorithm and state-space model estimation, running on a 2018 MacBook Pro 15" laptop with 2.9 GHz i9 processor, 32 GB RAM, with R 3.6.2.

Median distances between state-space model-estimated and interpolated GPS locations were within 8 km for all species and data types, with most species and data types having 95\% of estimated locations within 12 km of GPS locations (Table \ref{tab:dists}). Northern gannets were an exception, with 95-th percentiles extending $> 40$ km for all Argos data types (Table \ref{tab:dists}). Importantly, the median accuracy of state-space model-estimated locations, regardless of Argos data type, were all smaller or comparable to those of pre-filtered but un-modelled KS locations (Table \ref{tab:dists}). Across species, the weighted average ($\pm$ se) improvement of state-space model-estimated location accuracy relative to un-modelled KS location accuracy was: LS = 0.21 $\pm$ 0.60 km; KF = 0.14 $\pm$ 0.07 km; KS = 0.34 $\pm$ 0.05 km. 

Six of the 7 species' estimated tracks had median RMSD values under 5 km with all values under 10 km, regardless of Argos data type (Fig. \ref{fig:rmse}). Northern gannet tracks had considerably higher and more variable RMSD's (between 13 and 31 km), across all Argos data types (Fig. \ref{fig:rmse}). This is consistent with their considerably larger state-space model-predicted location uncertainty (Fig. \ref{fig:sppfits}). Both hawksbill turtle and southern elephant seal tracks had declining RMSD values as Argos data frequency and precision increased (Fig. \ref{fig:rmse}), and this was consistent with the increasing resolution and precision of their state-space model-predicted tracks (Fig. \ref{fig:sppfits}). Conversely, leopard seal and northern gannet tracks showed no such pattern, which was consistent with the general lack of increasing resolution of both the observed and predicted tracks (Fig. \ref{fig:sppfits}). Results were similar, although with overall lower RMSD values, when comparing state-space model estimated locations to the temporally closest GPS location within $\pm 10$ min (Fig. \ref{fig:Srmsd}).

\subsubsection*{Effect of $\psi$ parameter}
Inclusion of the $\psi$ parameter resulted in lower RMSD values, on average, implying that Argos error ellipses under-represent the true location uncertainty in the general north - south direction (Fig. \ref{fig:psi}). This result was less pronounced with fits to Argos Kalman smoother locations, with 81\% of individuals having a $\log \Delta$ RMSD $<$ 0 versus 90\% of individuals for Argos Kalman filter locations (KF $\Delta$ RMSD: median = -0.57 km, range = -3.78,0.45; KS $\Delta$ RMSD: median = -0.27 km, range = -3.34, 0.85). Of the four species, predicted locations for hawksbill turtle tracks were least likely to benefit from re-scaled error ellipses, with most individuals having $\log \Delta$ RMSD values close to or $>$ 0 (Fig. \ref{fig:psi}). It is unclear whether this is due to: 1) their relatively low absolute RMSD values (Fig. \ref{fig:rmse}); 2) their slightly more circular error ellipses (Table \ref{tab:dataerr}), where the $\psi$ re-scaling effect would be less pronounced; or, 3) a combination of the two. 

\subsubsection*{Argos KS accuracy}
Argos Kalman smoother locations were less accurate by an average of 0.34 km without subsequent state-space model filtering (Table \ref{tab:dists}; compare KS and pf\_KS values), although comparisons of $\log \Delta$ RMSD were variable both within and among species (Fig. \ref{fig:kfs}). The mean $\log \Delta$ RMSD across species implied a average 6\% increase in accuracy with subsequent state-space model filtering of Argos KS locations. However, results were equivocal for southern elephant seals and hawksbill turtle tracks were typically more accurate without any subsequent state-space filtering (Fig. \ref{fig:kfs}). 

\section*{Discussion}
We presented a continuous-time model for animal movement, fit in a state-space framework that allows flexible handling of Argos satellite telemetry data. The model was initially intended for automated quality control of large Argos animal tracking data sets, but is broadly applicable for any Argos location data. Using Argos - GPS double tagged animals, we assessed the accuracy of model-estimated locations, comparing across three types of Argos data where possible. Median accuracy was within 4 km for most species and data types, with state-space model-estimated locations being slightly more accurate (by 0.1 - 0.3 km on average) than the best quality CLS Kalman smoother locations. Median root mean squared deviations were typically at or under 5 km for 6 of the 7 species studied. In most cases, RMSD values were lowest when fitting to Argos Kalman smoother data and highest when fitting to Argos Least-Squares or Kalman filter data, although the within-species differences in RMSD between data types were typically small. Although the model was evaluated over a limited number of individuals and species, it is apparent that the accuracy and spatio-temporal resolution of inferred locations is situational.

Highlighting this situational aspect are the northern gannet results (Table \ref{tab:dists}; Figs. \ref{fig:sppfits} \& \ref{fig:rmse}), which are clearly distinct from the other species. Accuracy of model-estimated locations was approximately 4-5 times worse than for other species, although absolute magnitude is subject to the approach used for matching model-estimated and GPS locations (compare Figs. \ref{fig:rmse} \& \ref{fig:Srmsd}). Unlike other species where median distances between model-estimated and GPS locations either declined consistently or were similar when comparing LS to KF and KF to KS data types, gannets had the lowest median distances for fits to LS data and had far broader distributions of distance across the 3 data types. We suspect this pattern may arise from the considerably faster mean travel rates of northern gannets (12 km h\textsuperscript{-1}, with cruising speeds up to 45 km h\textsuperscript{-1}) compared to the other species (approximately 0.7 - 3 km h\textsuperscript{-1}). Similarly, Lopez et al. \cite{Lopez:2015} reported lower overall coverage probabilities of error ellipses estimated by their Kalman filter and Kalman smoother algorithms for two avian species analyzed in comparison to other platforms (terrestrial and marine mammals, sea turtles, ships and drifters). Combined, this implies that Argos error ellipses may be more strongly underestimated for species/platforms that travel faster and/or at higher altitude.

McClintock et al. \cite{McClintock:2015} used a bivariate $t$-distribution, parameterised by the Argos error ellipse information, to model location measurement error. Their estimates of the $t$ degrees of freedom parameter implied that the Argos error ellipses do not fully explain location measurement error. To avoid computational challenges associated with $t$-distribution parameter estimation, we used a two-step approach for dealing with location measurement error in Argos Kalman filter and Kalman smoother data. First, we identified and removed potentially large outliers using a travel-rate filter \cite{Freitas:2008} prior to fitting the state-space model, as per \cite{Johnson:2008,Patterson:2010}. Although underestimation of location error was acknowledged by Lopez et al. \cite{Lopez:2014,Lopez:2015} and has been reported by others \cite{Boyd:2013,McClintock:2015}, it is unclear why occasional, apparent hugely underestimated error ellipses are present in the Kalman filter and Kalman smoother data. Second, we accounted for potential Argos error ellipse underestimation by including the $\psi$ parameter to inflate the semi-minor axis. We adopted this approach given the observation that Argos error ellipses often have semi-minor axes vastly smaller than corresponding semi-major axes, resulting in ``squashed'' error ellipses (Figs. \ref{fig:SellipseHB} - \ref{fig:SellipseNG}). We found that in most cases the $\psi$ parameter contributed to more accurate location estimates, implying that the error ellipses commonly underestimate the true uncertainty in Argos-measured locations. This result is evident but less pronounced when fitting to Kalman smoother versus Kalman filter data. Location estimates were more accurate for at least some individuals of all species, however, hawksbill turtles and northern gannets appeared least likely to benefit from the $\psi$ re-scaling effect (see Fig. \ref{fig:psi}). Both of these species had somewhat more circular error ellipses, in comparison to the leopard and southern elephant seals, and thus any possible contribution of $\psi$ would be reduced. Ultimately, we are unsure why Argos error ellipses appear to be so commonly biased low in the semi-minor axis direction (generally north - south).

Where possible, both Kalman filter and Kalman smoother data types were included in this study. We found, in most cases, that the model-estimated locations were most accurate when using the Kalman smoother data, but on average by less than 200 m compared with fits to Kalman filter data. Although the Kalman smoother data should represent optimal estimates of location because information along the entire movement track is used to update and smooth each location estimate, we show that fitting the state-space model to these estimates can further improve location accuracy in some cases (by an average reduction in error of approximately 6\%). The Kalman smoother data are not provided in the default, near real-time service from CLS, rather they are only available with post-processing by CLS at an additional cost. There are two points to be made about this. First, the smoothing algorithm is a standard approach that can be implemented rapidly, with computing requirements no greater than the Kalman filter. It could be applied in near real-time. Second, a near real-time Kalman smoother would result in the best available location estimates changing as new data became available. This incremental improvement, due to information gain propagating backwards in time, would reduce as locations become less recent. This should be of little consequence to most wildlife users who typically do not use their data in near real-time, and users who do require near real-time data may see greater benefit in more accurate locations even if they are subject to change in retrospect.

Our state-space model produced location estimates with a median accuracy comparable to or greater than CLS' Kalman smoother locations, regardless of input Argos data type. This implies that users can obtain similar or better accuracy than CLS' Kalman Smoother locations by applying the state-space model to their Least-Squares or Kalman filter data. Therefore the method we describe is a viable alternative to the CLS' fee-based reprocessing service. The Laplace approximation approach employed in Template Model Builder models states (velocity and location) as unknown random effects, providing a most likely estimate of the current state from the posterior of it's location given all available data, both forward and backward in time. This is precisely what a Kalman smoother does. That our model can improve on the CLS Kalman smoother's location estimates may imply that uncertainty is somehow not well-propagated from the raw Doppler shift data available to CLS through to the location estimates available to users. If this is indeed the case, it is unclear why this is so. The issue may be due to necessary trade-offs between accuracy and precision versus providing a near real-time location service for a multitude of moving platforms, of which wildlife are a small component.

\subsection*{Spatio-temporal resolution and spatial accuracy}
It is important to note that when comparing GPS locations with those from models fitted to Argos-measured locations, accuracy is interlinked with the temporal resolution (sampling rate) of Argos relative to GPS locations. As GPS resolution is typically greater than Argos, comparisons to determine spatial accuracy of estimated locations are confounded by this difference. No model fit to Argos-measured locations alone can resolve all the nuances of a movement path that are present in higher resolution GPS data. This discrepancy will be reflected in measures of spatial accuracy, unless GPS data are suitably sub-sampled or interpolated. 

We interpolated GPS locations to the times of the Argos-measured locations to which the state-space model was fitted. Our reasoning was that interpolation of the generally higher resolution GPS data should be less corrupted by spatial error than a similar interpolation of the lower resolution and irregularly occurring model-estimated locations. Sub-sampling GPS locations by matching them with the temporally closest model-estimated location, commonly used elsewhere \cite{Costa:2010,Hoenner:2012,Lopez:2015}, resulted in lower RMSD or greater (apparent) accuracy than comparison with the linearly interpolated GPS locations. These lower RMSD values, however, were based on fewer (n $<$ 10) temporally matched pairs of model-estimated and GPS locations for some species/individuals (Fig. \ref{fig:Srmsd}); using a 20-min window. Although sample sizes could be increased by choosing a wider time window, the potential for biased comparisons would increase differently across species due to their different spatio-temporal scales of movement.

Fits to the three Argos location types from the same individuals showed that movement pathways can be predicted with increasing spatial resolution, i.e., resolve greater spatial detail despite the same prediction time interval (2 h), and precision as the number of Argos-measured locations increased (transition from Least-Squares to Kalman filter data) and as their uncertainty decreased (transition from Kalman filter to Kalman smoother data). One of the main advantages of Argos' Kalman filter over the older Least-Squares method is a gain in the number of location estimates, mostly by resolving locations from the single transmissions between tag and satellite that Least-Squares can not \cite{Lopez:2014}. This increase in resolution and precision is case-dependent, however, as species with lower overall proportions of class A and B locations do not gain as many new locations when transitioning from Least-Squares to Kalman filter data. This case-dependency is likely tied to typical surface time intervals of diving species, and, for those species spending the majority of time in air, on the magnitude of their travel rates. 

On different issue of scale, many ecological analyses of animal tracking data consider remotely sensed or other environmental data at spatial resolutions (2 - 10 km; e.g., \cite{Hindell:2020}) approaching the state-space model accuracy limits found here. This highlights the need for researchers to consider the appropriate resolution of their environmental data given their specific questions and the limitations of their location estimates. Fitting a state-space model to Argos tracking data is not a panacea. Researchers should consider carrying location uncertainty estimates provided by state-space models through to subsequent ecological analyses. For example, by repeatedly sampling from the location uncertainty, conducting the analysis, and pooling results (sensu \cite{McClintock:2017}). This can be done either completely through the whole analysis or partially via subsequent sensitivity analysis.

\section*{Conclusions}
The state-space model developed and validated here can be used to obtain quality-controlled animal locations from Argos Least-Squares or Kalman filter data in near real-time, with median accuracy comparable to or marginally better than CLS' reprocessed Kalman Smoother data. Our model also accounts for apparent north-south bias in Kalman filter- and Kalman smoother-derived error ellipses. 

The model's near real-time capability provides the best estimates of location, given the available data, that can be continually updated as new data arrive via the Argos system. This rapid, continual quality control of animal tracking data is necessary as near real-time monitoring and forecasting of ocean states increasingly incorporates oceanographic data from animal-borne sensors, and as the need for dynamic ocean management grows in our increasingly exploited and rapidly changing oceans.

Although the model was developed for fast, automated quality control processes, its simplicity and ease of use also make it suitable for manual use by researchers wishing to conduct quality control of historical or otherwise less immediate Argos data.


\begin{backmatter}

\section*{Competing interests}
  The authors declare that they have no competing interests.

\section*{Author's contributions}
    Conceived and designed the study: IDJ, CRM, TAP. Developed methodology: IDJ, TAP. Performed the analyses: IDJ. Contributed data: DPC, PDD, WJG, BJG, CG, XH, SK, PWR, SCV, SW, MJW, MAH, RGH, CRM. Wrote the paper: IDJ. Edited the paper: All.

\section*{Acknowledgements}
  We thank M Weise and B Woodward for motivating the validation study, H Lourie for assistance with CLS reprocessing, and M Holland and K Wilson for facilitating data access. IDJ supported by Macquarie University's co-Funded Fellowship Program and by external partners: Office of Naval Research grant N00014-18-1-2405; the Integrated Marine Observing System - Animal Tracking Facility; the Ocean Tracking Network; Taronga Conservation Society; Birds Canada; and Innovasea/Vemco. TAP supported by CSIRO Oceans \& Atmosphere internal research funding scheme. CG thanks the Institut Polaire Français Paul Emile Victor (IPEV programs 109, H.Weimerskirch and 1201, C.Gilbert) and Terres Australes et Antarctiques Françaises (TAAF) for logistical and field support. WJG and SCV thank Greg \& Lisa Morgan for field assistance and were funded by NERC New Investigators Grant (NE/G001014/1), the Peninsula Research Institute for Marine Renewable Energy and EU INTERREG Project CHARM III. DPC and PWR thank National Oceanographic Partnership Program, the Office of Naval Research, the Moore, Packard, and Sloan Foundations, and California Sea Grant Program. SSK supported by a National Science Foundation Office of Polar Projects research grant. XH and SW supported by the Australian Government under the Caring for Country Initiative, the Anindilyakwa Land Council, the Northern Territory Government, Charles Darwin University, and the ANZ Trustees Foundation – Holsworth Wildlife Research Endowment.


\bibliographystyle{vancouver} 
\bibliography{refs}      




\begin{figure}[!b]
\includegraphics[width=4.8in]{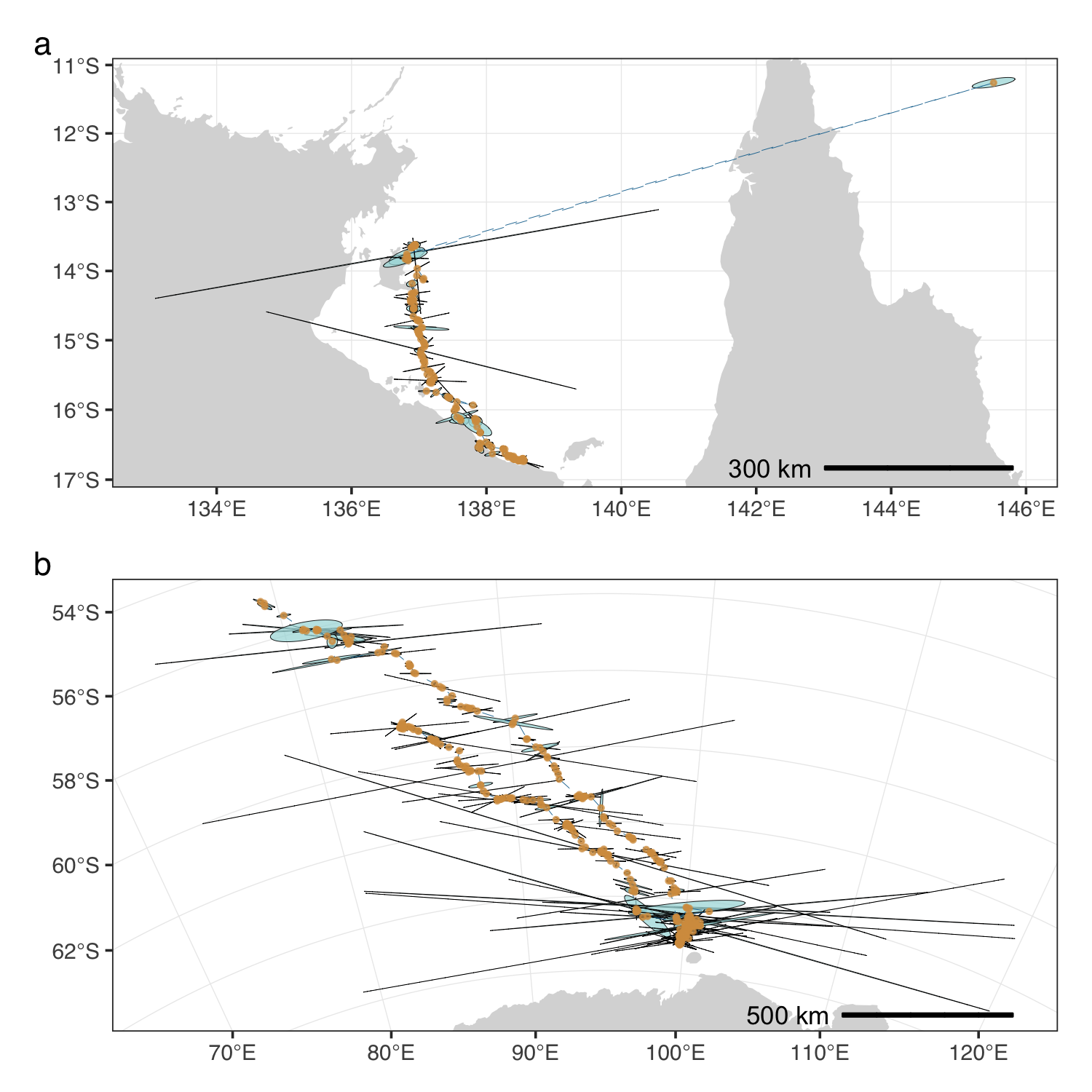}
 \caption{\csentence{Unfiltered Argos Kalman filter (KF) locations (gold points) and error ellipses (pale blue with black borders) for (a) hawksbill turtle and (b) southern elephant seal.} Locations are connected by dashed blue lines. Scale bars at lower right provide an indication of the magnitude of errors. The extreme outlier location at top right (a) is approximately 650 km away from the preceding and following locations, and has a vastly underestimated error ellipse. }\label{fig:EE}
\end{figure}
\afterpage{
\begin{landscape}
\begin{figure}[!p]
    \includegraphics[width=9in]{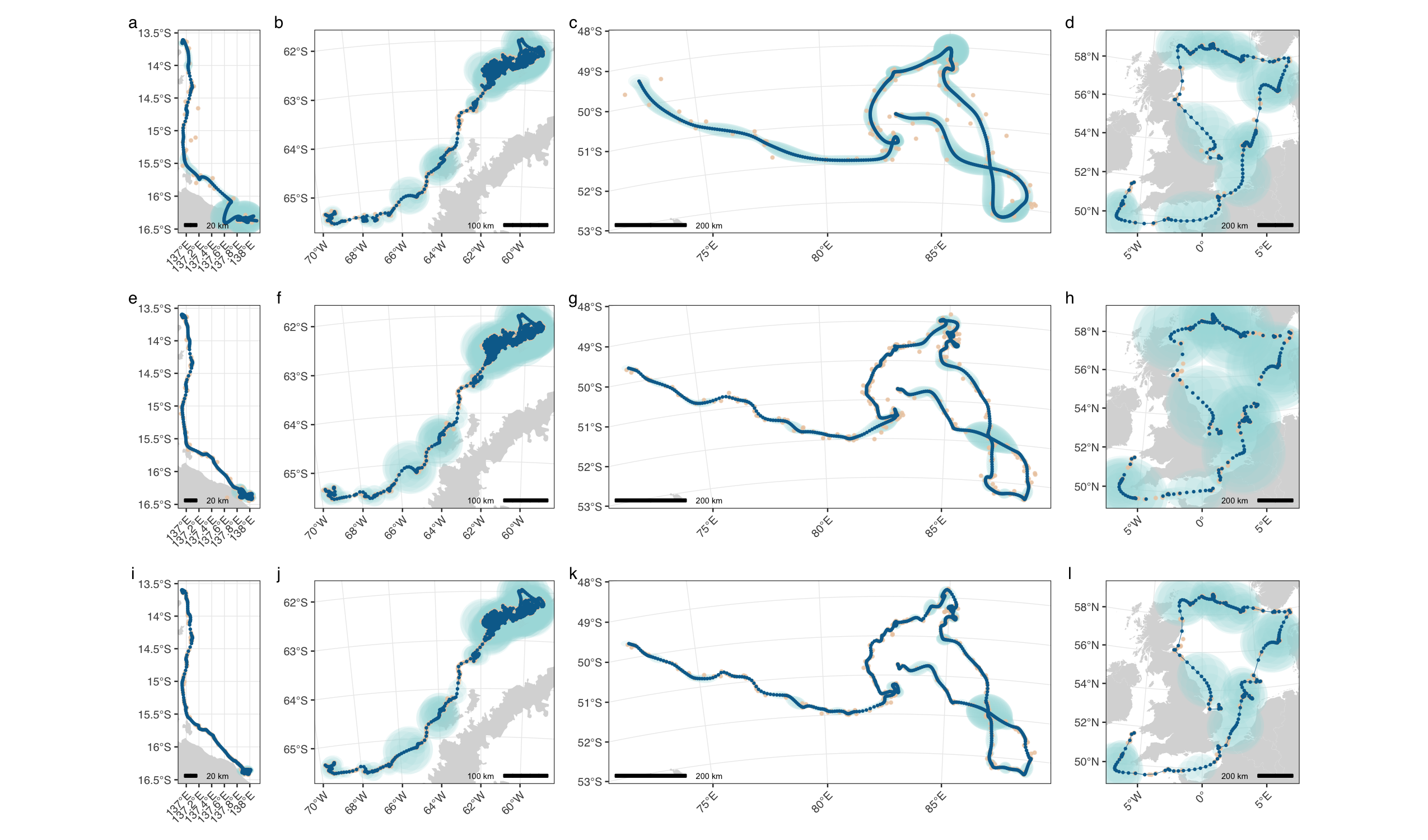}
  \caption{\csentence{State-space model fits to the three Argos satellite data types obtained from a hawksbill turtle (a,e,i), a leopard seal (b,f,j), a southern elephant seal (c,g,k), and a northern gannet (d,h,l).} State-space model-predicted locations (dark blue circles) smooth through the Argos data types (gold circles) - Least-Squares (a,b,c,d), Kalman filter (e,f,g,h) and Kalman smoother (i,j,k,l) - at regularly specified 2-h intervals. The 95\% confidence intervals on the predicted locations (light blue) are also displayed. Spatial scale in km is indicated at lower left or right of each panel.}\label{fig:sppfits}
      \end{figure}
\end{landscape}
\clearpage
}
 
\afterpage{      
\begin{figure}[!p]
\includegraphics[width=4.8in]{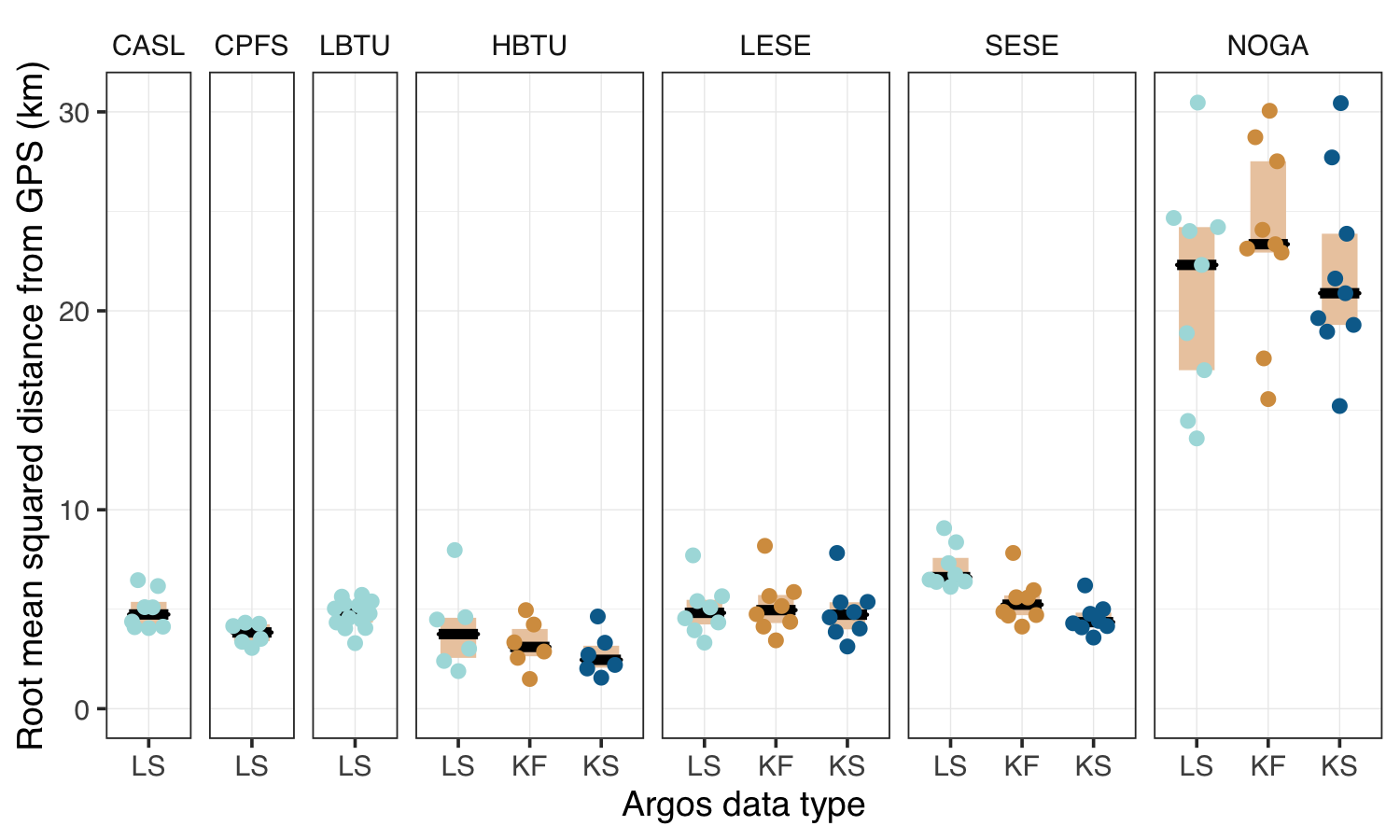}
  \caption{\csentence{Root mean squared distance (RMSD in km) between state-space model estimated locations and corresponding interpolated GPS locations by Argos data type and species.} Three species - California sea lion (CASL), Cape fur seal (CPFS), and leatherback turtle (LBTU) - only had Argos least-squares (LS) data available, the remainder - hawksbill turtle (HBTU), leopard seal (LESE), southern elephant seal (SESE), and northern gannet (NOGA) - had data that were reprocessed by CLS Argos to yield all 3 Argos data types - LS, Kalman filter (KF) and Kalman smoother (KS). Individual RMSD values (filled circles), inner quartile range (beige box) and medians (black bar) are displayed.}\label{fig:rmse}
      \end{figure}
\clearpage
}      

\afterpage{
\begin{figure}[!p]
\includegraphics[width=4.8in]{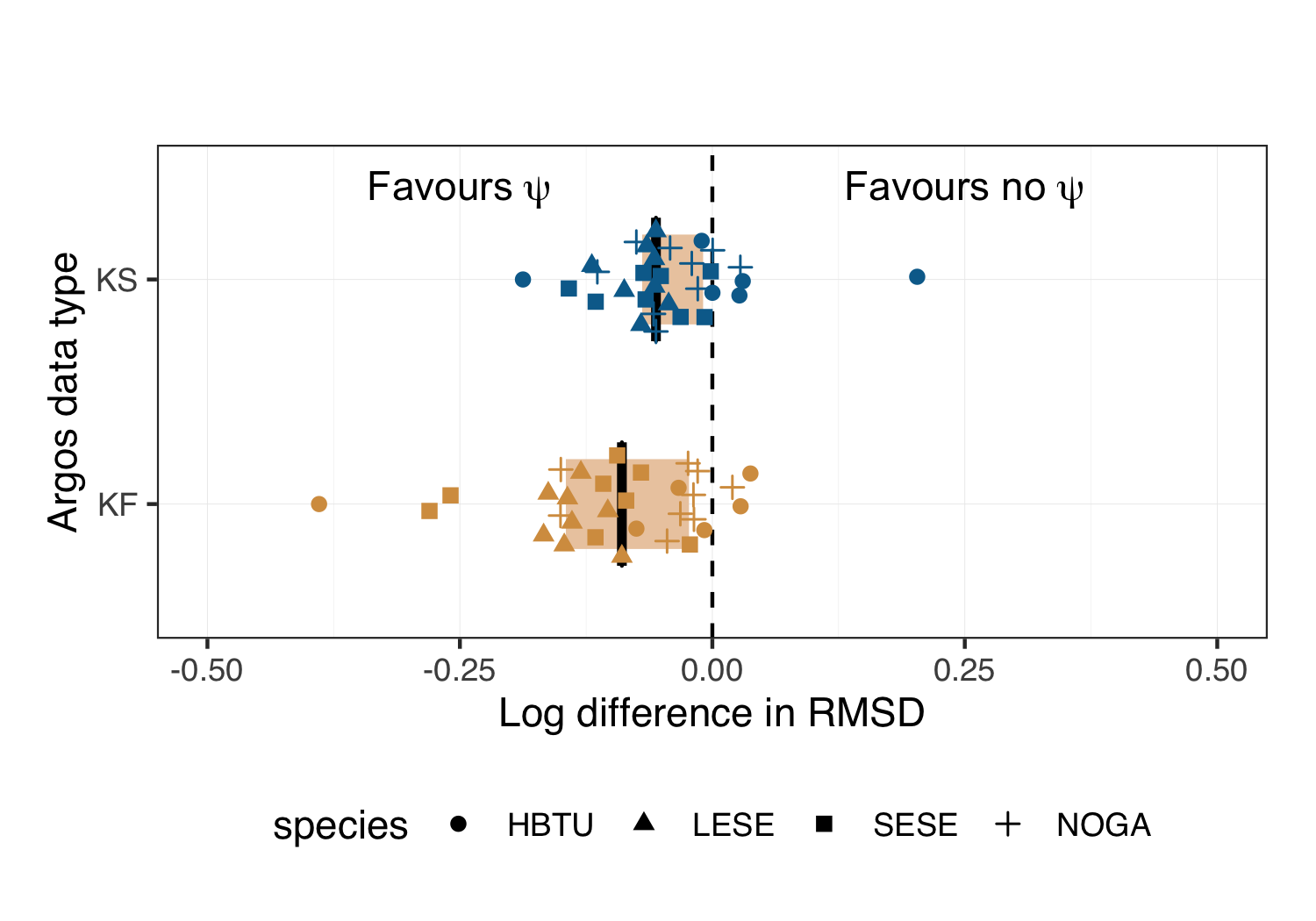}
  \caption{\csentence{Accuracy of estimated locations from models fit with and without a $\psi$ parameter to re-scale the Argos error ellipse semi-minor axis length.} The log difference in root mean squared distance ($\log \Delta$ RMSD) between state-space model-estimated locations and interpolated GPS locations was calculated for the two models fit to each individual animal. Negative values indicate the model with a $\psi$ parameter is more accurate, whereas positive values indicated the model without a $\psi$ parameter is more accurate. Results are pooled across species with individual log differences in RMSD displayed as species-specific shapes (gold, blue). The inner quartile range (beige box) and medians (black bar) are displayed.}\label{fig:psi}
      \end{figure}

\begin{figure}[!p]
\includegraphics[width=4.8in]{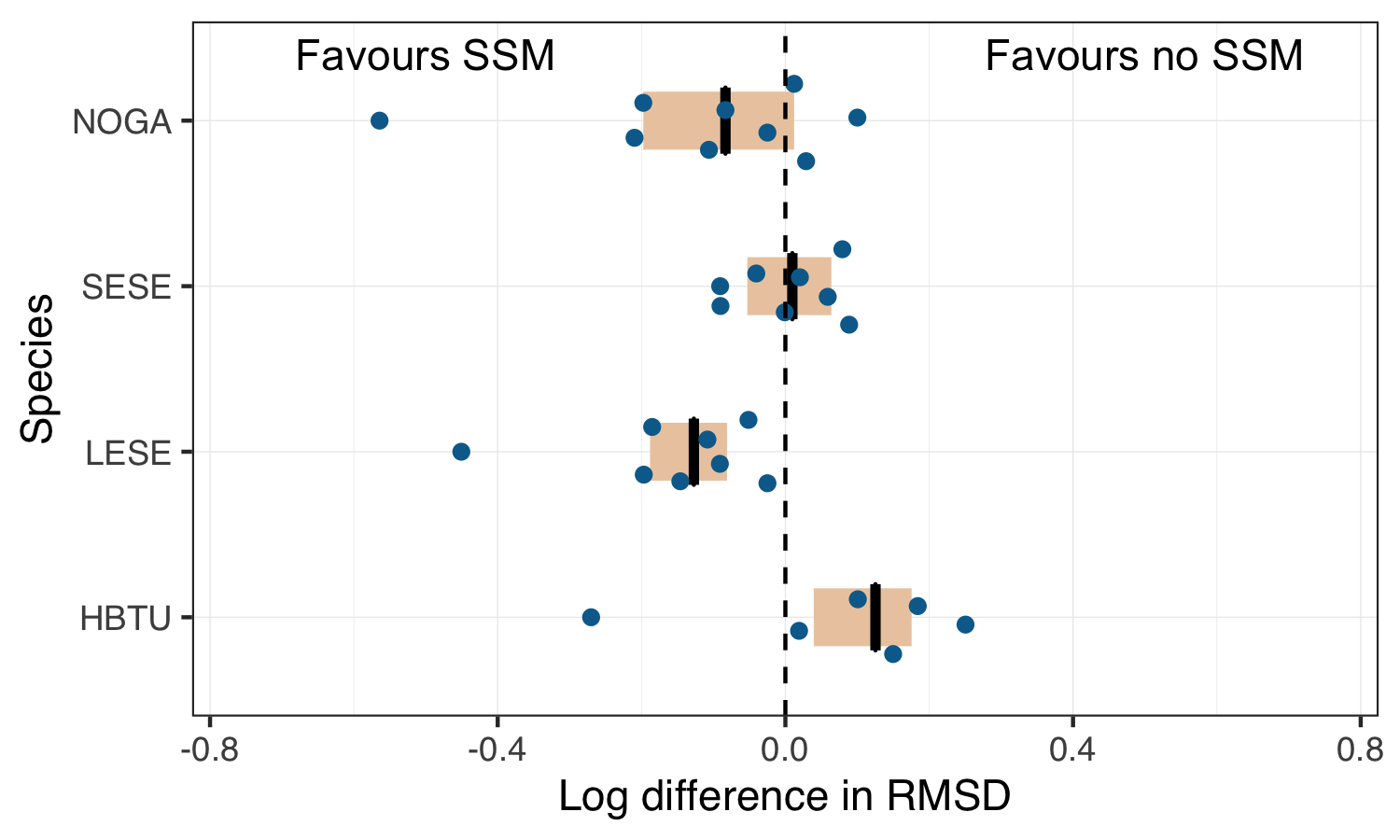}
\caption{\csentence{Accuracy of Argos Kalman smoother locations with and without subsequent state-space model filtering.} The log difference in root mean squared distance ($\log \Delta$ RMSD) between locations and interpolated GPS locations was calculated for each individual animal. In both cases, the Kalman smoother locations were subjected to the same travel-rate filtering to remove highly implausible observations. Negative values indicate the state-space model estimated locations are more accurate. The inner quartile range (beige box) and medians (black bar) are displayed.}\label{fig:kfs}
\end{figure}
\clearpage
}



\afterpage{
\begin{landscape}
\begin{table}[!p]
\caption{Number of individual data sets by species and data type. Argos data types are: Least-Squares (LS); Kalman filter (KF); Kalman smoother (KS). Mean track durations were calculated from the data after removing periods of prolonged data gaps. Tag programming details were not available for all deployments, so Argos and GPS sampling rates were calculated from the unfiltered data.}\label{tab:sppdat}
\centering
\begin{tabular}[t]{rccccrrrrcc}
\toprule
\multirow{2}{*}{Species} & \multirow{2}{*}{Common name} & \multirow{2}{*}{Code} & Deployment & Mean track & \multicolumn{4}{c}{Data type} & GPS sample & Fastloc\\
\cmidrule(l{3pt}r{3pt}){6-9}
 &  &  & year(s) & duration (d) & LS & KF & KS & GPS & rate (min) & GPS\\
\midrule
\emph{Zalophus californianus} & California sea lion & CASL & 2007 & 79 & 8 & . & . & 8 & 58 & Y\\
\emph{Arctocephalus pusillus} & Cape fur seal & CPFS & 2007 & 25 & 6 & . & . & 6 & 49 & Y\\
\emph{Dermochelys coriacea} & leatherback turtle & LBTU & 2008/12 & 92 & 13 & . & . & 13 & 195 & Y\\
\emph{Eretmochelys imbricata} & hawksbill turtle & HBTU & 2009/10 & 24 & 6 & 6 & 6 & 6 & 124 & Y\\
\emph{Hydrurga leptonyx} & leopard seal & LESE & 2018 & 171 & 8 & 8 & 8 & 8 & 47 & Y\\
\emph{Mirounga leonina} & southern elephant seal & SESE & 2009/11/12/14 & 53 & 11 & 11 & 11 & 11 & 28 & Y\\
\emph{Morus bassanus} & northern gannet & NOGA & 2010 & 23 & 9 & 9 & 9 & 9 & 60 & N\\
\bottomrule
\end{tabular}
\end{table}
\end{landscape}
\clearpage
}

\afterpage{
\begin{table}[!p]
\caption{Argos track summary statistics by species and data type. prop'n A,B is the proportion of all locations that are in quality class A and B. Error ellipse shape is the ratio of semi-minor to semi-major axis length, with values closer to 1 indicating a more circular shape. Shape and area statistics were calculated from values pooled among individuals within species and Argos data type.}\label{tab:dataerr}
\centering
\begin{tabular}[t]{rcrrrrr}
\toprule
\multicolumn{1}{c}{Species } & \multicolumn{1}{c}{Argos} & \multicolumn{1}{c}{total n} & \multicolumn{1}{c}{prop'n} & \multicolumn{1}{c}{Error ellipse} & \multicolumn{2}{c}{Error ellipse area (km)} \\
\cmidrule(l{3pt}r{3pt}){6-7}
code & type & locations & A,B & shape & median & 95\textsuperscript{th} \%-ile\\
\midrule
HBTU & LS & 603 & 0.71 & . & . & .\\
HBTU & KF & 1 038 & 0.84 & 0.24 & 17.31 & 120.29\\
HBTU & KS & 1 038 & 0.84 & 0.27 & 8.44 & 60.31\\
\addlinespace
LESE & LS & 32 780 & 0.47 & . & . & .\\
LESE & KF & 41 367 & 0.57 & 0.13 & 1.73 & 69.23\\
LESE & KS & 41 367 & 0.57 & 0.14 & 1.03 & 31.50\\
\addlinespace
SESE & LS & 3 152 & 0.82 & . & . & .\\
SESE & KF & 5 016 & 0.91 & 0.13 & 34.40 & 780.11\\
SESE & KS & 5 016 & 0.91 & 0.14 & 16.26 & 338.18\\
\addlinespace
NOGA & LS & 1 568 & 0.36 & . & . & .\\
NOGA & KF & 2 066 & 0.52 & 0.18 & 25.43 & 5 573.62\\
NOGA & KS & 2 066 & 0.52 & 0.19 & 16.73 & 2 104.87\\
\bottomrule
\end{tabular}
\end{table}

\begin{table}[!p]
\caption{Accuracy in km of state-space model-estimated locations and pre-filtered KS locations (pf\_KS), by species and Argos data type. The pf\_KS locations had the pre-filter algorithm applied but not the state-space model. Median and 95\textsuperscript{th} percentile statistics were calculated from distances to GPS locations, pooled among individuals within species and Argos data type.}\label{tab:dists}
\centering
\begin{tabular}[t]{rrrrrrrrr}
\toprule
\multicolumn{1}{c}{ } & \multicolumn{2}{c}{LS} & \multicolumn{2}{c}{KF} & \multicolumn{2}{c}{KS} & \multicolumn{2}{c}{pf\_KS} \\
\cmidrule(l{3pt}r{3pt}){2-3} \cmidrule(l{3pt}r{3pt}){4-5} \cmidrule(l{3pt}r{3pt}){6-7} \cmidrule(l{3pt}r{3pt}){8-9}
Species & median & 95\textsuperscript{th} \%-ile & median & 95\textsuperscript{th} \%-ile & median & 95\textsuperscript{th} \%-ile & median & 95\textsuperscript{th} \%-ile\\
\midrule
CASL & 1.70 & 10.15 & . & . & . & . & . & .\\
CPFS & 1.87 & 8.03 & . & . & . & . & . & .\\
LBTU & 2.77 & 10.06 & . & . & . & . & . & .\\
HBTU & 1.60 & 8.11 & 1.69 & 7.08 & 1.47 & 6.05 & 1.62 & 5.91\\
LESE & 1.88 & 11.11 & 2.05 & 11.61 & 1.89 & 10.90 & 2.25 & 12.62\\
SESE & 4.40 & 15.03 & 3.24 & 12.06 & 2.97 & 9.50 & 3.29 & 11.02\\
NOGA & 6.04 & 43.95 & 7.50 & 52.62 & 7.37 & 46.47 & 7.85 & 53.67\\
\bottomrule
\end{tabular}
\end{table}
\clearpage
}

\afterpage{
\section*{Additional Files}
\subsection*{Additional file 1 --- Global map of Argos least-squares tracking data by species.}

 \begin{figure}[!h]
 \renewcommand\thefigure{S1}
  \includegraphics[width=4.8in]{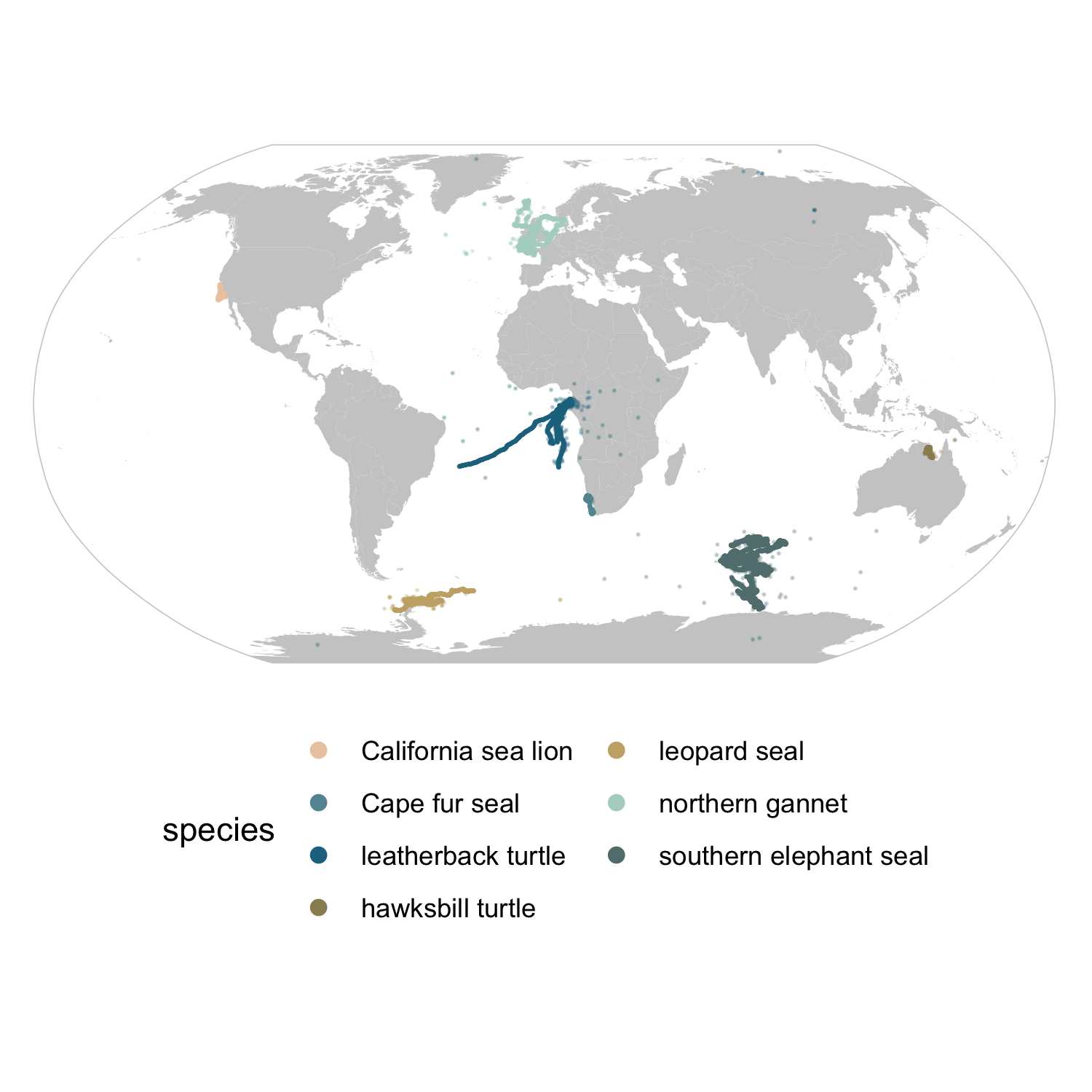}
  \caption{\csentence{Argos tracking data by species.} Locations displayed are unfiltered Argos Least-Squares data.
      }\label{fig:sppmap}
      \end{figure}
\clearpage
}

\afterpage{
  \subsection*{Additional file 2 --- Argos Kalman filter and Kalman smoother error ellipses along a hawksbill turtle track.}
\begin{figure}[!h]
    \renewcommand\thefigure{S2}
    \includegraphics[width=4.8in]{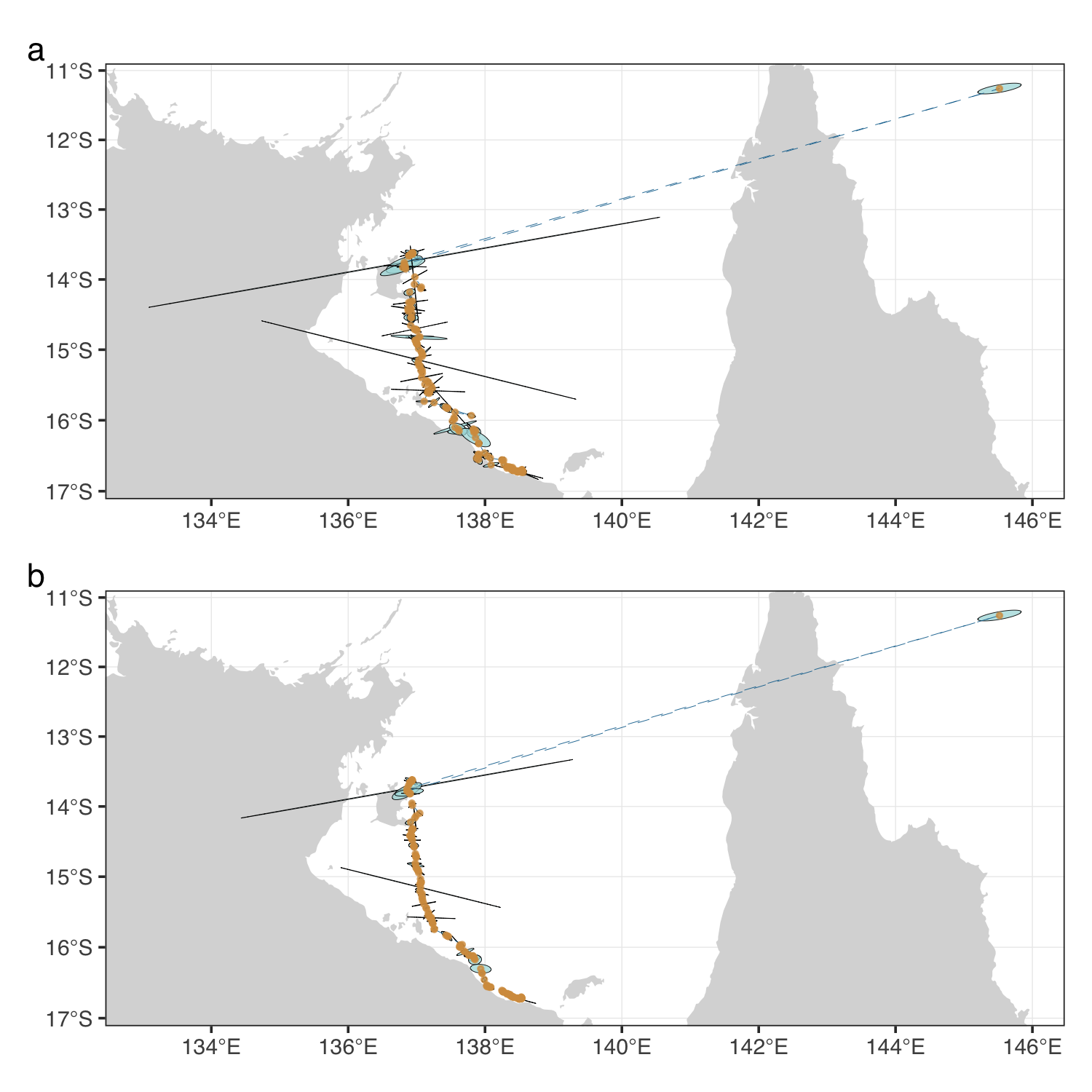}
 \caption{\csentence{Argos error ellipses for a representative hawksbill turtle track.} The maps display (a) Argos Kalman filter- and (b) Argos Kalman smoother-measured locations (beige) and error ellipses (light blue with black borders). Locations are connected by dashed blue lines.}\label{fig:SellipseHB}
\end{figure}

    The majority of visible ellipses are highly compressed on their semi-minor axes for both Argos data types. The dashed blue track line helps highlight the outlier location (upper right) that clearly has a vastly under-estimated error ellipse. Application of the Kalman smoother (b) did not reduce this outlier or improve its uncertainty estimate.
\clearpage
}

\afterpage{
  \subsection*{Additional file 3 --- Argos Kalman filter and Kalman smoother error ellipses along a leopard seal track. }
\begin{figure}[!h]
    \renewcommand\thefigure{S3}
    \includegraphics[width=4.8in]{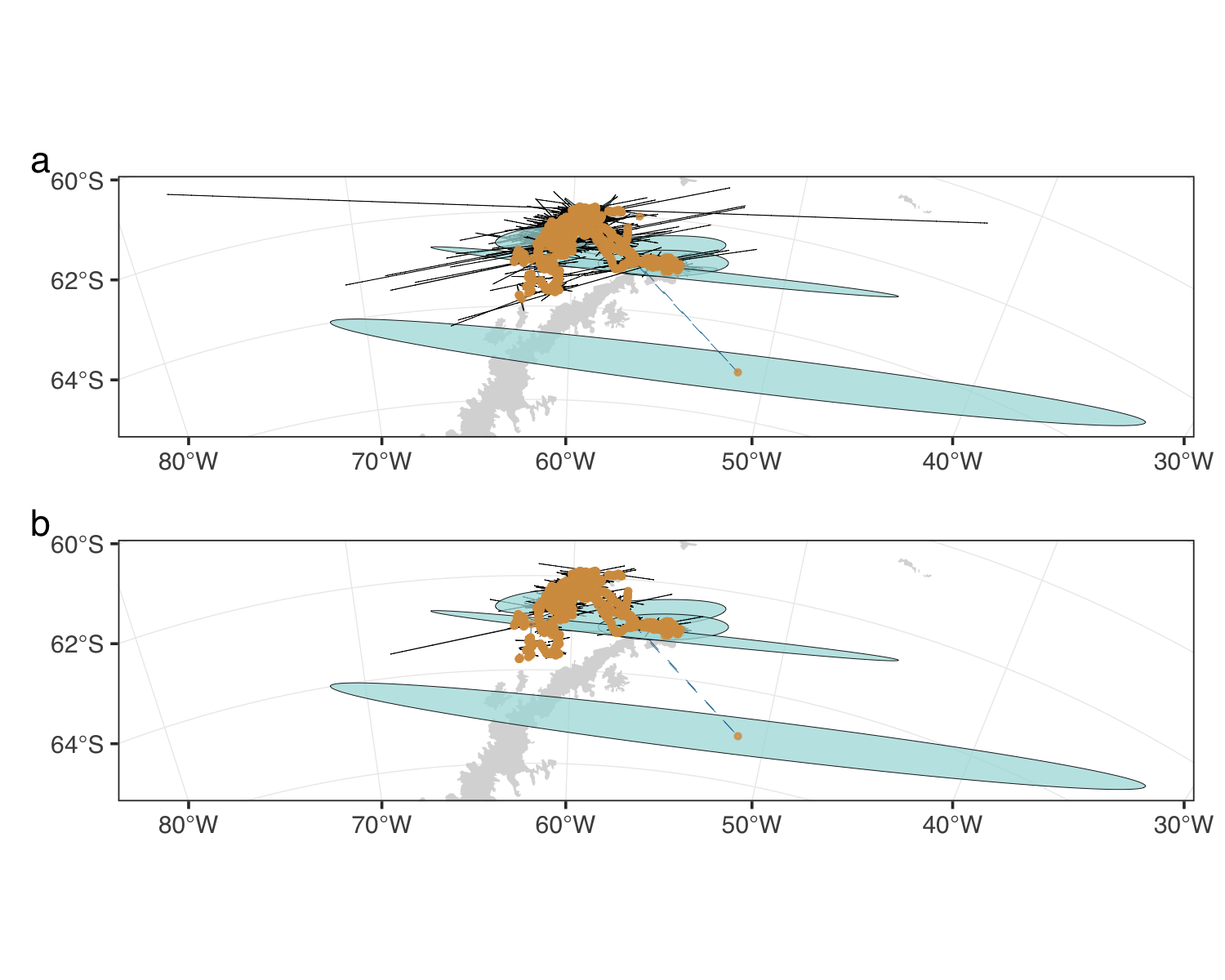}
 \caption{\csentence{Argos error ellipses for a representative leopard seal track.} The maps display (a) Argos Kalman filter- and (b) Argos Kalman smoother-measured locations (beige) and error ellipses (light blue with black borders). Locations are connected by dashed blue lines.}\label{fig:SellipseLS}
\end{figure}

    The majority of ellipses are highly compressed on their semi-minor axes for both Argos data types. The dashed blue track line helps highlight the outlier location (lower right) that has an error ellipse with under-estimated semi-minor axis. Application of the Kalman smoother (b) did not reduce this outlier or improve its uncertainty estimate.
\clearpage
}

\afterpage{
  \subsection*{Additional file 4 --- Argos Kalman filter and Kalman smoother error ellipses along a northern gannet track. }
\begin{figure}[!h]
    \renewcommand\thefigure{S4}
    \includegraphics[width=4.8in]{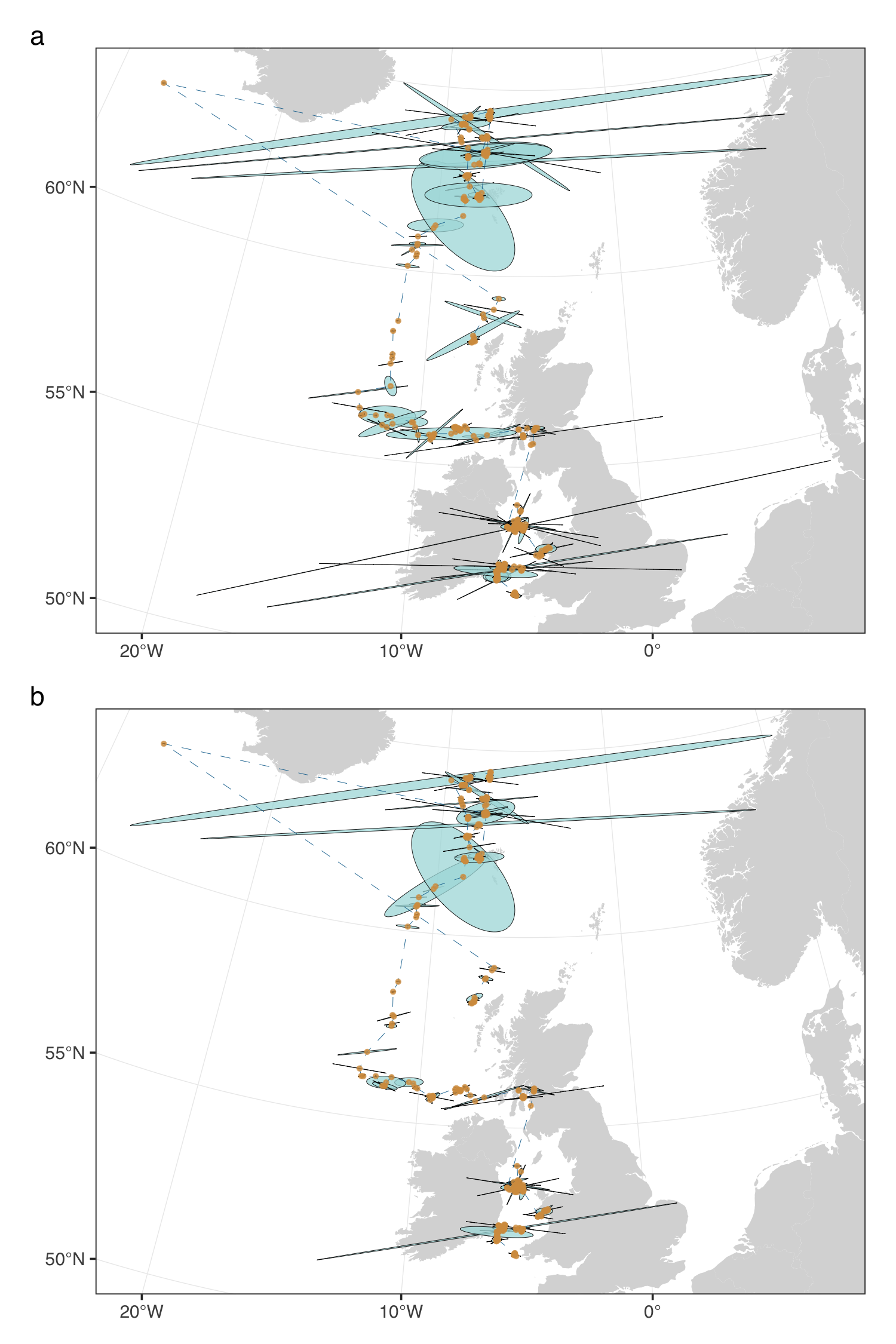}
 \caption{\csentence{Argos error ellipses for a representative northern gannet track.} The maps display (a) Argos Kalman filter- and (b) Argos Kalman smoother-measured locations (beige) and error ellipses (light blue with black borders). Locations are connected by dashed blue lines.}\label{fig:SellipseNG}
\end{figure}

\clearpage
}

\afterpage{
  \subsection*{Additional file 5 --- Alternate validation using temporally closest GPS location with $\pm 10$ min of state-space model-estimated locations.}
  
\begin{figure}[!h]
\renewcommand\thefigure{S5}
\includegraphics[width=4.8in]{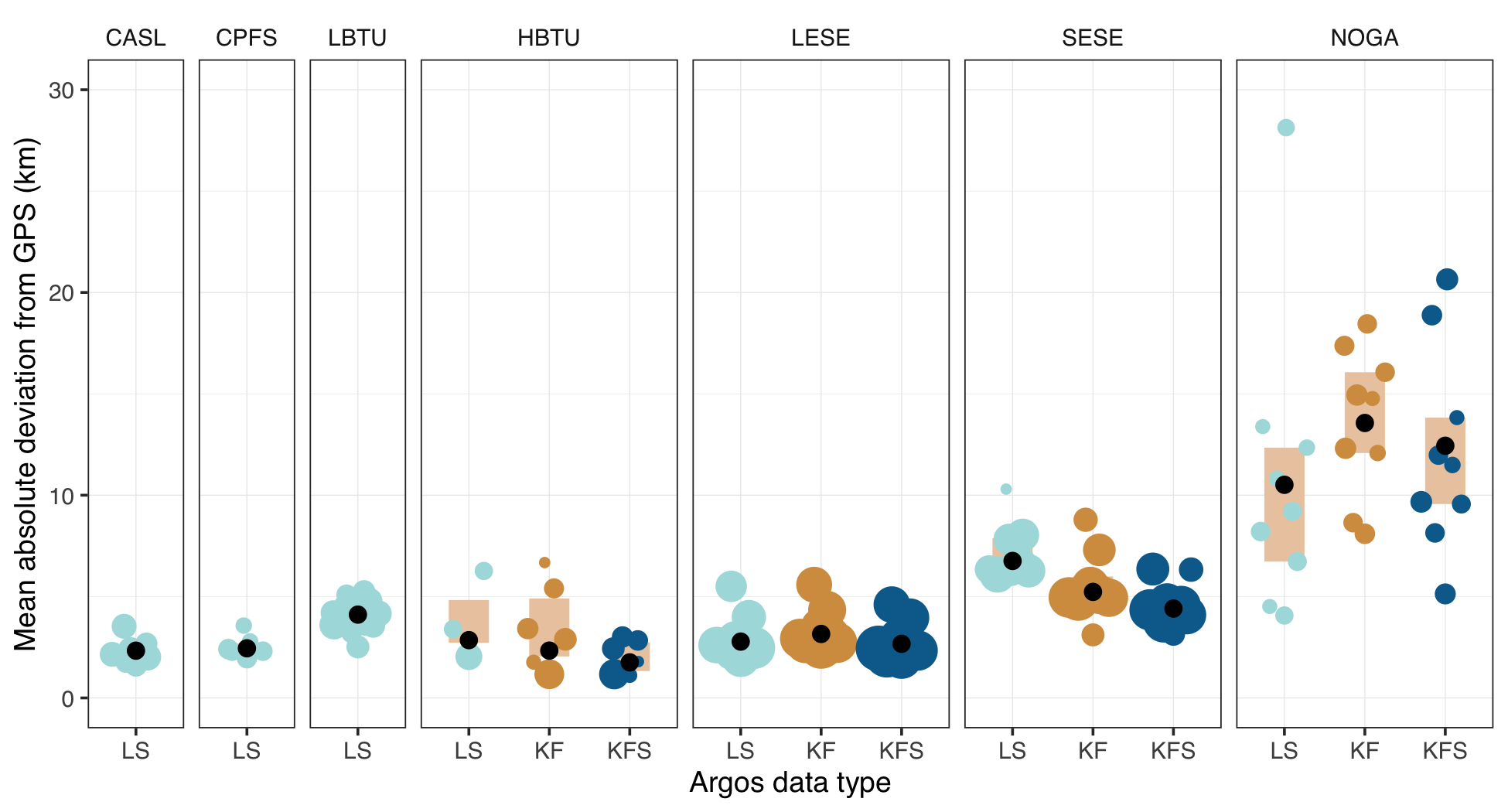}
 \caption{\csentence{Root mean squared distance (RMSD in km) between state-space model estimated locations and their corresponding temporally closest GPS locations within $\pm 10$ min, by Argos data type and species.} Size of RMSD points (coloured) is proportional to $\sqrt{n}$ distances between temporally matched pairs of model-estimated and GPS locations. Black points are $n$-weighted means by species and data type.}\label{fig:Srmsd}
\end{figure}
  
\clearpage
}

\end{backmatter}
\end{document}